\documentclass[a4paper]{article}
\usepackage[english]{babel}
\usepackage[utf8x]{inputenc}
\usepackage{amsmath}
\usepackage{amsfonts}
\usepackage[numbers,round]{natbib}
\usepackage{graphicx}
\usepackage[colorinlistoftodos]{todonotes}
\usepackage[affil-it]{authblk}

\title{Bridging scales in cancer progression: Mapping genotype to phenotype using neural networks}
\author{Philip Gerlee\footnote{Corresponding authors: philip.gerlee@moffitt.org and alexander.anderson@moffitt.org}, Eunjung Kim \& Alexander R.A. Anderson$^*$}
\affil[]{Department of Integrated Mathematical Oncology}
\affil[]{H. Lee Moffitt Cancer Center and Research Institute}
\affil[]{12902 Magnolia Drive Tampa, FL 33612, USA}
\begin{document} 
\maketitle

\begin{abstract}
In this review we summarize our recent efforts in trying to understand the role of heterogeneity in cancer progression by using neural networks to characterise different aspects of the mapping from a cancer cells genotype and environment to its phenotype. Our central premise is that cancer is an evolving system subject to mutation and selection, and the primary conduit for these processes to occur is the cancer cell whose behaviour is regulated on multiple biological scales. The selection pressure is mainly driven by the microenvironment that the tumour is growing in and this acts directly upon the cell phenotype. In turn, the phenotype is driven by the intracellular pathways that are regulated by the genotype. Integrating all of these processes is a massive undertaking and requires bridging many biological scales (i.e. genotype, pathway, phenotype and environment) that we will only scratch the surface of in this review.  We will focus on models that use neural networks as a means of connecting these different biological scales, since they allow us to easily create heterogeneity for selection to act upon and importantly this heterogeneity can be implemented at different biological scales. More specifically, we consider three different neural networks that bridge different aspects of these scales and the dialogue with the micro-environment, (i) the impact of the micro-environment on evolutionary dynamics, (ii) the mapping from genotype to phenotype under drug-induced perturbations and (iii) pathway activity in both normal and cancer cells under different micro-environmental conditions.
\end{abstract}

\section*{Introduction}
There has been a great deal of effort focused on trying to define molecular signatures of many different aspects of cancer (including risk, progression and treatment response) with limited success. This lack of success is largely due to the multiscale aspect of cancer \cite{Anderson2008}, and was recently brought into sharp focus with the landmark publication by Swanton and colleagues \cite{Gerlinger:2012kx} in which they showed that a single tumour can contain multiple distinct genetic signatures. This so called, intratumour heterogeneity, is at the heart of our lack of success with such biomarkers and emphasizes a need for a deeper understanding of how genetic changes alter intracellular signalling and how this signalling alters the cellular phenotype. To complicate matters further, all of this intrinsic heterogeneity does not happen in a vacuum nor does it appear instantaneously, rather it occurs in a temporally dynamic spatially heterogeneous micro-environment that directly impacts tumour heterogeneity. 

{ Another critical aspect of this heterogeneity is the role that it plays in therapuetic escape. The recent advent of targeted therapies, where drugs are employed to target specific genetic mutants, whilst appearing as initially a huge success for cancer control, have also been hampered by heterogeneity and given way to drug resistance (see \cite{Ellis:2009vn} for a review).  Drug resistance, in general, is a major problem in cancer treatment and therefore understanding how this resistance develops and can be slowed or even prevented is a current major focus of the cancer community (see \cite{Holohan:2013zr} for a review). A recent strategy to deal with this resistance is to utilize combination therapies, that combine targeted drugs with more broad base chemotherapuetic agents (see \cite{Saunders:2012ys} for a review) and has achieved mixed results. What is becoming clear is that a deeper understanding of how selection (via drugs or otherwise) alters heterogeneity at the phenotypic scale and differentially at the geneotypic scale, will allow us to develop smarter more effective therapies with already available drugs.}

In the last decades the network paradigm has become ubiquitous in cancer biology \cite{Kreeger2010}, and the idea that a biological system can be viewed as a conglomeration of discrete units that interact with a limited set of neighbouring units, has been applied on all scales of the disease, ranging from the regulation of RNA-molecules \cite{Lieberman2013}, the dynamics of intra-cellular signalling pathways \cite{Craene2013}, the interactions of different cell types within a tumour \cite{Basanta2013}, to the insight that metastatic spread occurs on a network defined by the circulatory system \cite{Scott2012}.

Network dynamics often run counter to linear and uni-directional reasoning, and hinder our understanding of the relation between mutations and phenotypic change.
Such nonlinear and counter-intuitive effects are known to arise already on the scale of intra-cellular signalling, where the inhibition of signalling molecules often has unexpected effects on signal transduction \cite{Cokol2011}, but the consequences are even greater when the micro-environment and tumour evolution are taken into account. Since evolution selects  phenotypes, not genotypes, tumours often consist of many distinct genotypes with similar phenotypes. The environment provides the key context upon which this evolutionary game is played out, but it is not a static environment. There is a direct feedback between the environment and successful phenotypes simply because these phenotypes can actively alter the microenvironment leading to changes in the selection pressure -- ultimately pushing the tumor along the path of progression.

Understanding the relationship between signalling, the micro-environment and tumour evolution
is required if we are to understand and rationally design effective anti-cancer therapies. In fact, even describing the unperturbed mapping from genotype to cellular phenotype is far beyond our current state of understanding. In a majority of cases our intuition is insufficient to grasp the many levels of simultaneous interactions occuring in a growing tumour, and hence it seems as if mathematical modelling with its ability to describe feedback and non-linearities as well as incorporate these dynamics on different scales, is the only way forward \cite{Byrne2010}. 


In this review we will discuss some recent mathematical modelling that has taken a few formative first steps in trying to understand this dynamic multiscale feedback system. We will focus specifically on individual-based models that use neural networks as a means of connecting these different biological scales of tumour growth.

\subsection*{Neural networks in cancer modelling}
Two different flavours of individual-based models of tumour growth have served as sources of inspiration for our neural network approach: evolutionary models in which a collection of static phenotypes grow and compete, and models with elaborate intra-cellular signalling, but without an evolutionary component. The former, exemplified by Anderson's \cite{Anderson2005} investigation of the impact of oxygen concentration and extra-cellular matrix density on tumour evolution, provided the backbone for modelling evolutionary dynamics, while models from the latter category, such as the work of Zhang et al. \cite{Zhang2007}, which investigated the role of EGFR-signalling on proliferation and migration, provided the inspiration for mapping micro-environment to cellular phenotype.

\begin{figure}
   \centerline{\includegraphics[width=13cm]{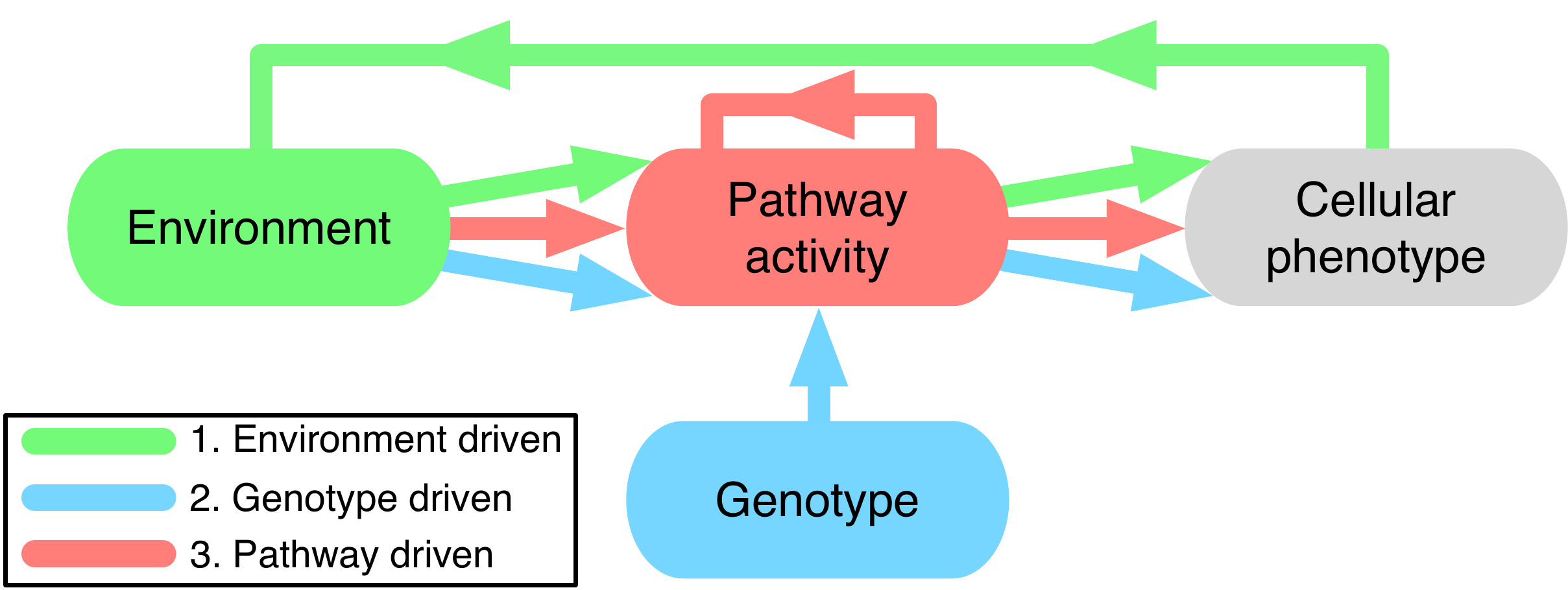}}
\caption{\label{fig:schematic}Colour coded summary of the different scales that each modelling approaches considers in this review. Coloured arrows emphasize where approaches overlap and where they are distinct.}
\end{figure}

The coalescence of these approaches in cancer modelling: selection and evolution of fixed phenotypes, and phenotypes that depend in a non-trivial way on environment and genotype, can be realised by letting the behaviour of each cell be determined by a neural network whose output -- the phenotype -- depends on the environment, genotype and pathway activity. In this capacity, neural networks serve as models of both the current state of cancer cell pathway activity (and hence phenotype) and changes due to mutation and selection. They also allow us to easily create heterogeneity for selection to act upon and importantly  this heterogeneity can be implemented at different biological scales (i.e. genotype, pathway, phenotype or environment). We will describe three models that focus on different aspects of this mapping (see fig.\ \ref{fig:schematic}): (i) the impact of the micro-environment on phenotypic evolutionary dynamics, (ii) the mapping from genotype to phenotype under drug-induced perturbations and (iii) pathway activity in both normal and cancer cells under different micro-environmental conditions.

\section*{1.\ The impact of the micro-environment on tumour evolution}
Artificial neural networks have traditionally been used for machine-learning tasks such as classification and prediction, some examples being the detection of heart abnormalities \citep{heart},  finger-print recognition \citep{finger} and breast cancer prediction \citep{breast}. In these settings the network is trained for detection with a data set that consists of a number of variables from each sample together with the outcome. 
There are two approaches to solve this problem, either by using a single network that is optimised with respect to the training set using an error minimising algorithm like back-propagation \citep{neural} or by using an evolutionary algorithm \citep{EANN}. 
Another application, more in line with the use in individual-based models, is the implementation of neural networks in evolving robot controllers \citep{robot}. Here the input to the network are sensors of the robot (e.g. proximity sensors) and the output of the network controls the motors. 

The use of neural networks as a means to map the micro-environment to tumour cell phenotypes  within a multi-cellular context was pioneered by Gerlee \& Anderson in a series of papers that investigated the impact of the micro-environment on tumour growth and evolution \cite{Gerlee2007,Gerlee2007a,Gerlee2009,Gerlee2010}. 

The approach had been hinted at by Bray \cite{bray1}, stating that \textit{"...systems of interacting proteins act as neural networks trained by evolution to respond appropriately to patterns of extracellular stimuli"} \citep{bray2}, and it was subsequently used in a single-cell context by Vohradsky as a means to model the $\lambda$-bacteriophage lysis/lysogeny decision circuit \cite{vohr}. In the context of cancer, a more data-driven application was developed by Naranyan et al., who used neural networks to model  temporal micro-array data \cite{nara}. In a similar vein Nelander et al.\ \cite{Nelander2008} used a feed-forward neural network to model the pathway activity in a breast cancer cell line under combinatorial  drug perturbation experiments.

Importantly these efforts were focused on the dynamics of a single cell, and the contribution of Gerlee \& Anderson was to incorporate this framework into an individual-based dynamic model, that allowed for mutation and hence evolution of the mapping from environment to phenotype to occur in a spatially meaningful manner. 

This approach has also found other applications in cancer modelling, e.g.\ as a means to model the evolutionary dynamics during drug treatment \cite{Mamun2013}. In that study, the dynamics of the model were shown to agree with \textit{in vitro} experiments, and also suggested novel mechanisms of action of the drug Maspin. The methodology has also been used for modelling the growth dynamics of multi-cellular spheroids in the presence of bioreductive drugs \cite{Kazmi2012}. That study focused on the problem of drug penetration and it was shown that the model could account for the growth rate, morphology and drug distribution seen in experiments. 
	
\subsection*{The environment-phenotype network}
\begin{figure}
   \centerline{\includegraphics[width=15cm]{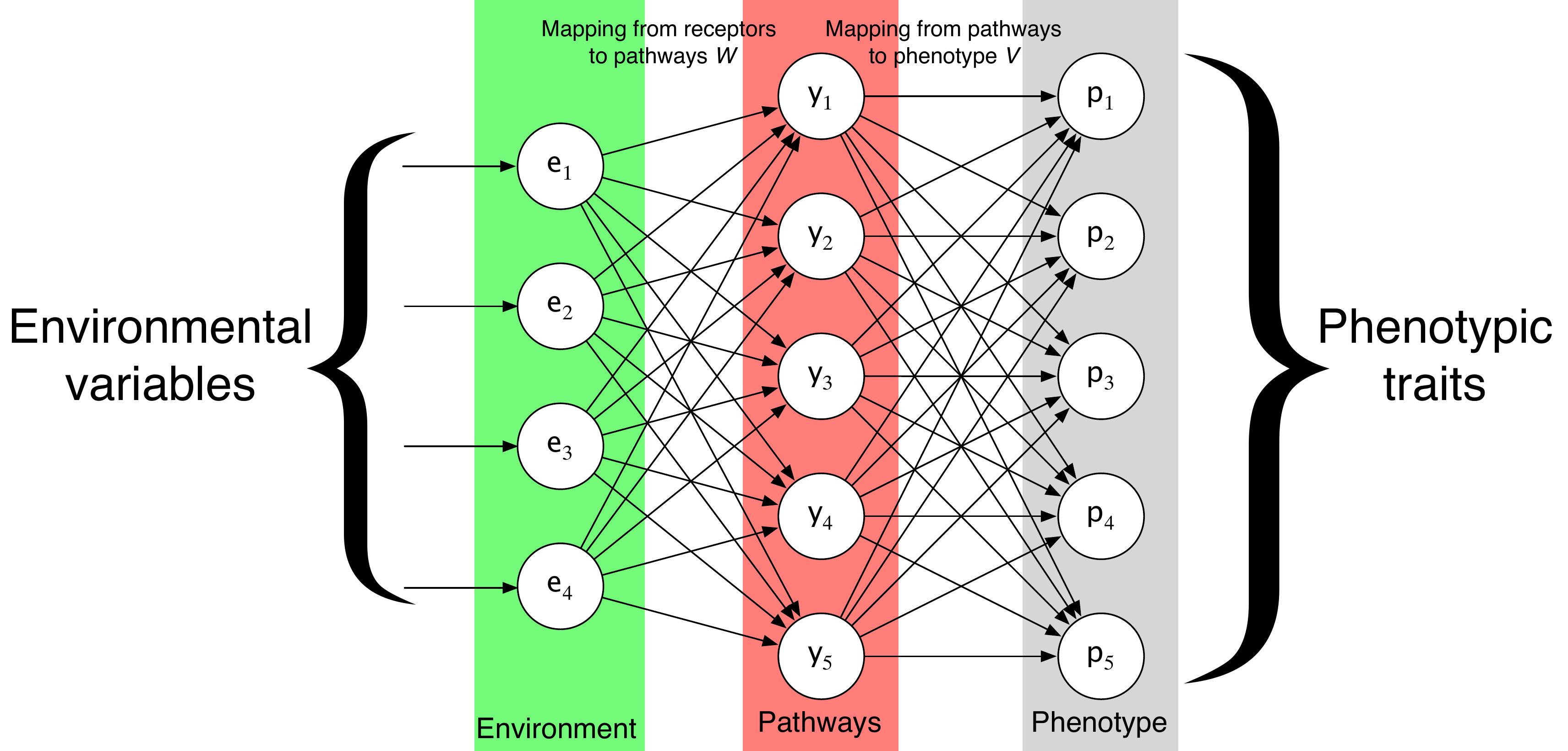}}
\caption{\label{fig:cellnet} { The structure of the response network.} Environmental variables (green) are presented to the input-layer and then fed through the pathways (red) where a phenotype is calculated (grey). Connections between these layers represent the mappings from receptors to pathways ($\mathbf{W}$) and from pathways to phenotype ($\mathbf{V}$).}
\end{figure}

The underlying idea behind the framework is the observation that the environment together with the signalling pathways of the cell jointly determines the behaviour or phenotype of a cell. Or from a more mechanistic viewpoint, that environmental stimuli fed through the  signalling network determined by the genotype gives rise to the phenotype. This provides the motivation for using feed-forward neural networks, that have a directional and layered structure, consisting of an input, hidden and output layer of nodes (see fig.\ \ref{fig:cellnet}). 

The output of the network is determined by feeding the micro-environmental variables to the input layer of the network and sequentially updating first the hidden layer and then the output layer. The output of the network dictates the phenotype of the cell and is deterministic, being dependent on the weight matrices ($\mathbf{W}$ and $\mathbf{V}$) and the environmental input (for details we refer to previous publications \cite{Gerlee2007,Gerlee2010}).

In order to capture the evolutionary dynamics of tumour growth this model includes mutations to the weight matrices. The mutations alter the connection strength between the nodes, which in turn changes how the cells respond to the micro-environment. By embedding these networks in individual cells and allowing them to mutate as they divide, we can create a caricature of an evolving tumour that responds to and modifies its micro-environment in a spatio-temporal manner. We achieved this by utilising a hybrid cellular automaton framework in which each cells behaviour  is driven by the neural network.

\subsection*{Hybrid model}
The hybrid model uses a cellular automata based approach that considers cells as simple  points on a grid, { that also contains continuous concentration fields of micro-environment factors, and together}  represent a thin slice of tissue. Each cell contains a feed-forward neural network that links environment to phenotype. {The initial matrices were parametrised in such a way that the behaviour of the cells would resemble that of early stage neoplastic cells (see Appendix for details).} The grid itself represents the micro-environment and can contain any number of possible variables, however, for simplicity we focus on the case where  the concentration of oxygen is the only micro-environmental factor considered. The dynamics of the oxygen concentration in space and time are controlled by, a discretised version of, the following partial differential equation
\begin{equation}\label{eq:oxyeq}
\frac{{\partial c(\vec x,t)}}{{\partial t}} = D_c \nabla^2 c(\vec x,t) - f_c (\vec x,t), 
\end{equation}
where $D_c$ is the diffusion constant of oxygen and $f_c(\vec x,t)$ gives the individual cell oxygen consumption rate for the cell at position $\vec x$ and time $t$. We make use of a fixed boundary condition to simulate a situation where the domain is surrounded by vessels.

In this hybrid model, cells perform three basic functions: proliferation, quiescence, and apoptosis (the phenotypic traits). The neural network within the cells controls these functions. The environmental variables (oxygen concentration and cell crowding, i.e.\ available space on the grid) are considered as input to the network. Please note that the node connections can vary and evolve from cell to cell, because of mutations that occur during cell division. The input is therefore processed by the cells’ individual networks in slightly different ways. For instance, a low value of oxygen concentration may trigger apoptosis in one cell, but not another, because their respective networks have evolved apart. Similar effects occur for quiescence and proliferation. Therefore, the model captures a crucial component of cancer progression, the generation of phenotypic variation within a population of growing cancer cells and how the selection imposed by different environmental conditions acts on this variation. More details on the precise implementation of this hybrid cellular automata model can be found in \cite{Gerlee2009b,Gerlee2009}.

\subsection*{The impact of oxygen on tumour growth and evolution}

The oxygen levels in a tissue are known to be a key regulator of tumour growth and progression, as it can both facilitate and inhibit the tumour in counter intuitive ways. We have previously used our neural network driven modelling paradigm to investigate the impact of varying oxygenation concentration on tumour growth and evolution \cite{Gerlee2009b,Gerlee2009} and summarise some key results below. 

\begin{figure}
   \centerline{\includegraphics[width=13cm]{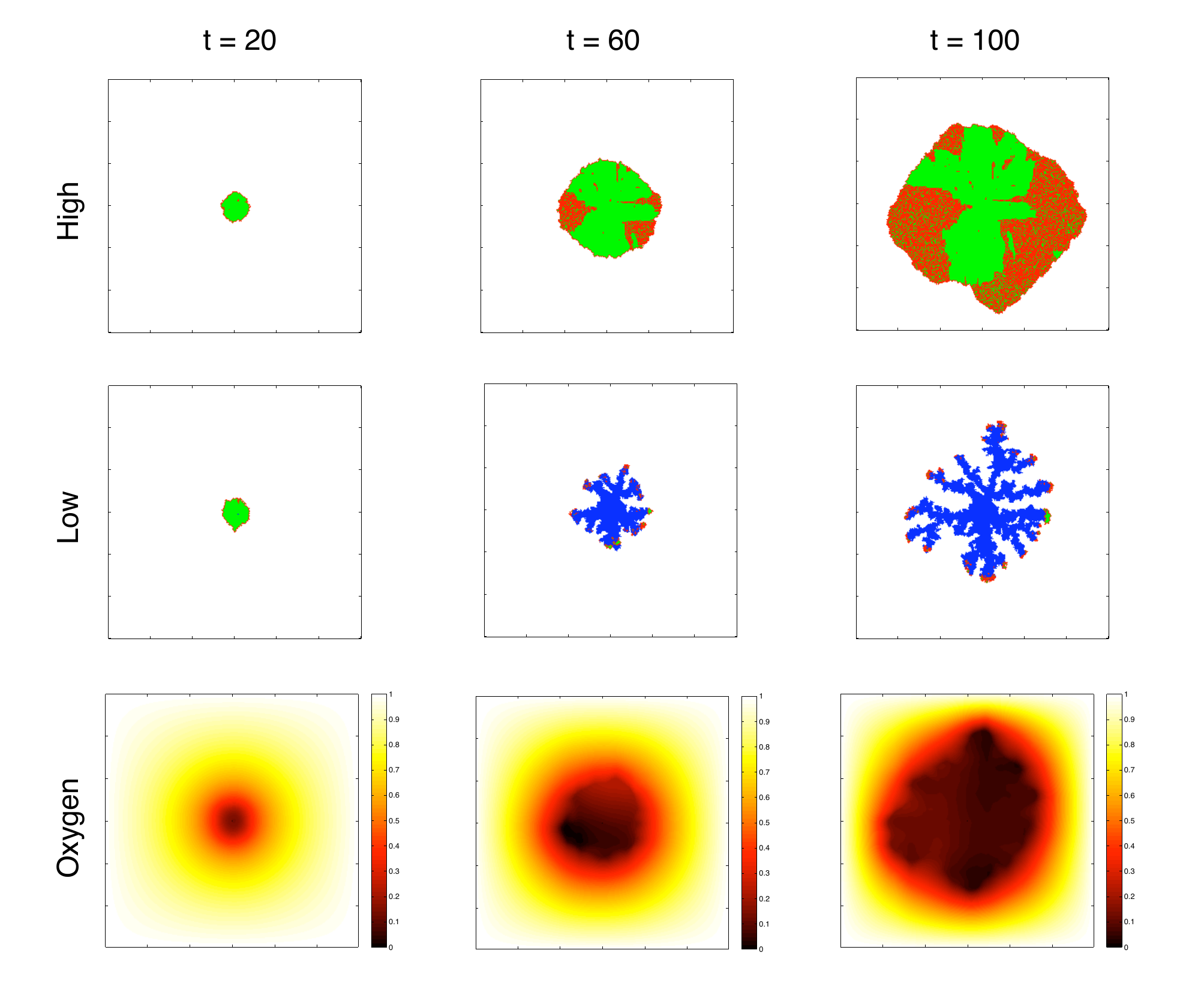}}
\caption{\label{fig:tumour}The two upper rows show the spatial distribution of cells at $t=$20, 60 and 100 days. Proliferating cells are coloured red, quiescent green and dead cells are blue. We can see a clear difference in the morphology of the tumours in the low and high oxygen case. The lower row shows the oxygen concentration in the low oxygen case.}
\end{figure}

Figure \ref{fig:tumour} shows a simulated tumour in different stages of growth at $t=20,60$ and 100 days for  two different values of the background oxygen concentration as well as the oxygen distribution at each time point for the low oxygen case. This figure clearly shows that a limited oxygen supply significantly  influences the growth dynamics and morphology of the tumour. In the high oxygen case the tumour only consists of proliferating (red) and quiescent (green) cells growing with a more compact and rounded morphology, while in the low oxygen case the tumour consists mostly of dead cells (blue) with proliferating cells at the tip of a more skeletal fingered structure. This structure is induced by competition for the limited oxygen, displayed in the lower panel and shows that a gradient of oxygen appears early in the simulation (t=20). The limited supply of oxygen means that the oxygen level rapidly drops below the apoptotic threshold in the centre of the tumour, which leads to the development of a necrotic core and subsequent invasive fingers. 

In this modelling framework, the phenotype of a cell is a function of its immediate micro-environment (oxygen concentration and local cell density), and therefore context dependent. In order to quantify the evolutionary change, the average phenotype was calculated as the proportion of the two-dimensional input space (oxygen concentration and number of neighbouring cells) that corresponds to each phenotypic response (proliferation, quiescence, apoptosis). This measure is summarised in a 3-dimensional vector $S$, that was termed the average phenotype or response vector, which reflects the behaviour of the cell. 
The initial phenotype, used as a seed in the simulation, had a measure of $S=($0.67, 0.18, 0.15$)$, which means that 67 \% of the input space corresponds to proliferation, 18 \% to quiescence and 15 \% to apoptosis (see figure \ref{fig:cancernet} in the Appendix for the actual network used). 

\begin{figure}
   \centerline{\includegraphics[width=16cm]{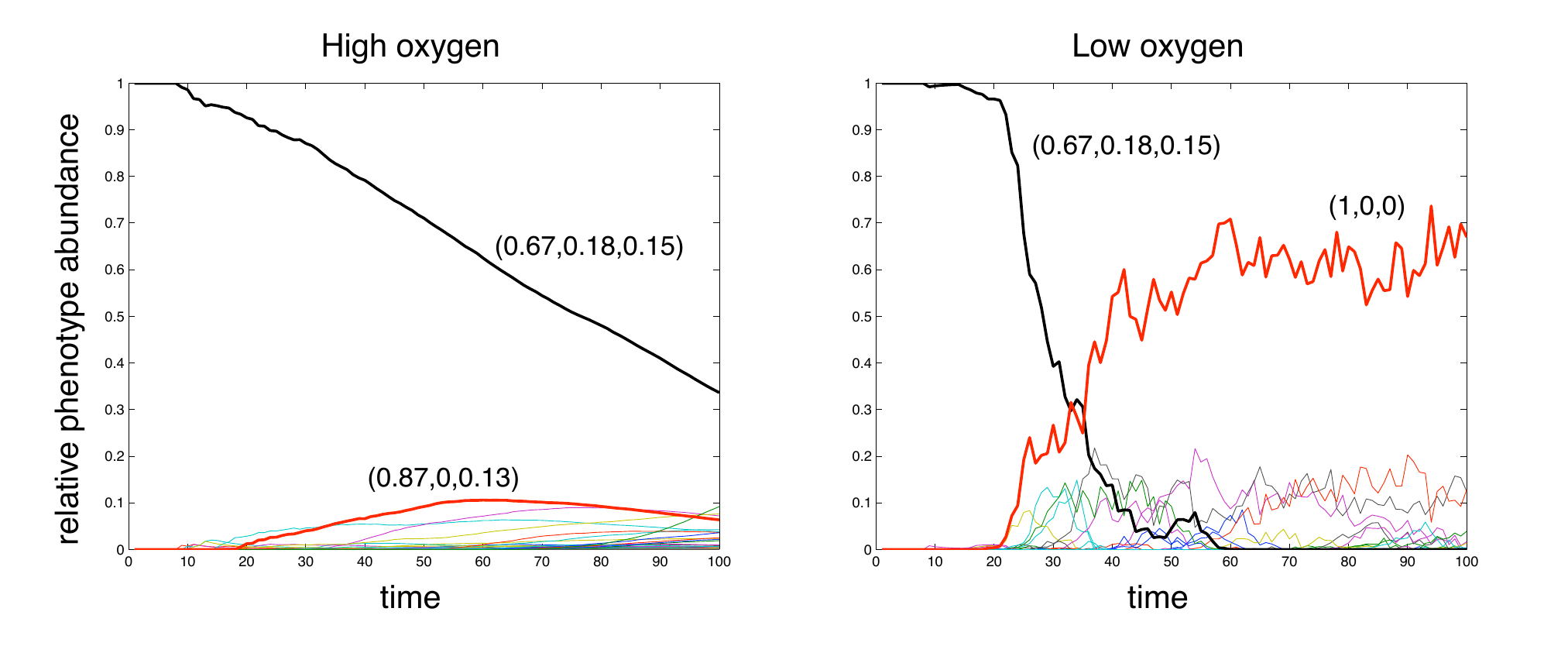}}
\caption{\label{fig:phenotype}The time evolution of the average phenotypes in the population for the high and low oxygen case. The most abundant phenotypes have been highlighted and their response vectors are displayed. {The fluctuations observed in the low oxygen case are a result of the smaller population size.}}
\end{figure}

This measure was used to analyse the evolution of phenotypes in the population by tracking the abundance of the average phenotypes in the population. Note that this is different from measuring the abundance of different subclones in the population as two subclones with distinct networks may still give rise to the same average phenotype. The time evolution of the phenotypes was measured for both oxygen cases and is shown in figure \ref{fig:phenotype}, where each line corresponds to a unique average phenotype and the most significant phenotypes have been highlighted. In the high oxygen environment we observe a steady decline of the initial phenotype and the emergence of several new phenotypes all with low abundances. This is in contrast with the dynamics in the low oxygen case where the initial phenotype decreases more rapidly and the population becomes dominated by a single phenotype.

{Similar results, both in terms of the impact of the microenvironment on the morphology and on the evolutionary dynamics, have been observed with other models. For example Anderson et al.\cite{anderson2} have shown that a harsh microenvironment (in terms of high extra-cellular matrix density and low oxygen concentration) gives rise the selection of a single subclone with aggressive traits. That model also recapitulates the fingered morphology, which has also been seen in a number of other types of models, e.g.\ level-set models \cite{Macklin2007}, hybrid cellular automata \cite{Ferreira2002} and cellular potts models \cite{Poplawski2009}.}

The phenotypic diversity is only one part of the story, as it represents the output of the neural network for each cell in the context of a given environment. Another important aspect of this result, is how the mappings changed as result of mutation and selection, i.e.\ how did the network connectivity of the cancer cells change? If we consider the overall connectivity of a subclone to represent its genotype, then we were able to show that the genotypic diversity was higher in the low compared to high oxygen case \cite{Gerlee2007}, suggesting that harsh selective growth conditions might give rise to evolutionary dynamics where the phenotypes in the population converge, while the genotypes diverge. This insight was only made possible through the use of a mutable mapping from environment and genotype to the phenotype. Clearly this is a gross simplification of the relationship between genotype and phenotype and we endeavour to refine this in the following section.

\section*{2. The emergence of resistance in a multi-drug context}
Understanding how the phenotype of a cell emerges from its genotype is, even in the absence of complicating environmental stimuli, a challenging problem, and even a partial understanding of this process could be helpful when designing anti-cancer therapies. This is particularly relevant now that targeted drugs, that inhibit specific genes or proteins, are becoming the standard of care for many types of cancer \cite{Sawyers2004}. If two targeted drugs are administered in combination, how do their effects on intra-cellular signalling interact, and equally important, how do they influence the evolution of resistance? Is the efficacy higher when the drugs are administered in parallel or in sequence, and in which case is the risk of resistance minimised?

The answers naturally depend on the architecture of the signalling network and the location of the targeted genes within that network. However, it is possible that some general principles exist, and we will give an example here of how neural networks can be used to tackle this problem.

\subsection*{The genotype-phenotype map}
Again we utilize neural networks to model the intra-cellular dynamics, but here we have focused on the effect of the genotype on intra-cellular signalling and disregard any environmental influence (see fig.\ \ref{fig:schematic}). Instead of viewing the genotype implicitly, as the weights and connections in the network, we now explicitly model it by letting the input nodes represent the mutational status of a subset of genes, that contribute to pathway activity, which then drive the phenotype $p$ of the cell. The application of a specific drug can then be modelled by modifications on different levels in the network, e.g.\ by removing the link between a gene and a pathway it contributes to. 

\begin{figure}[!htb]
\begin{center}
\includegraphics[width=13cm]{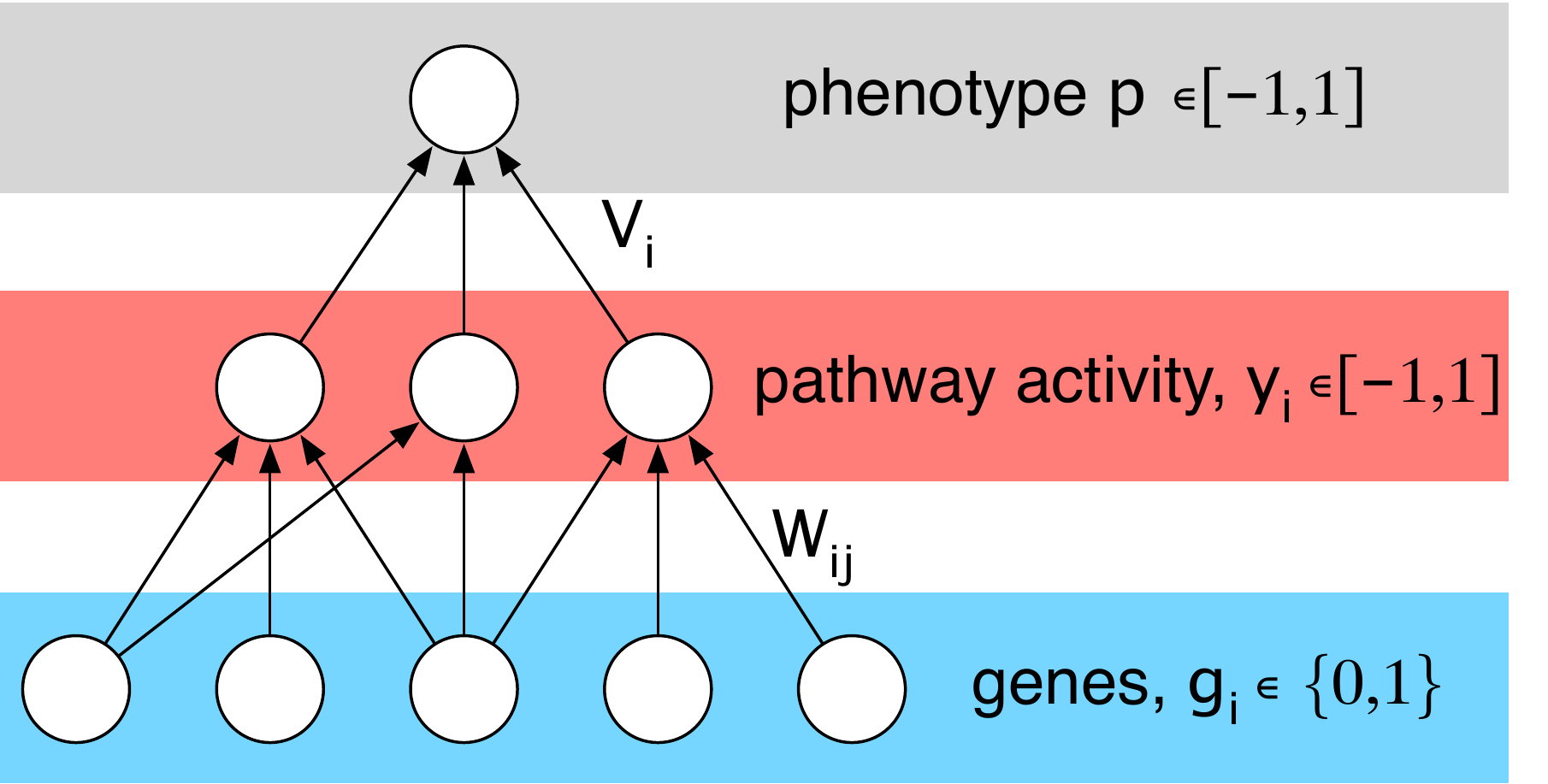}
\caption{\label{fig:fig1}Schematic of the genotype-phenotype map. In the model a number of genes contribute additively to distinct pathways that in turn influence the phenotype of the cell. Mutations at the genetic level, that activate/deactivate genes, propagate through the network and alter the phenotype.} 
\end{center}
\end{figure}

We model the influence of each gene on the phenotype of a subclone using a feed-forward neural network (see fig.\ \ref{fig:fig1}), and specify the fitness or growth rate of a subclone $i$ as a function of the phenotype $F_i=F(p_i)$, where we assume that an intermediate phenotype fitness optimal and define 
\begin{equation}
F(p) = e^{-(p-0.5)^2/\sigma}
\end{equation}
where $\sigma = 0.1$ determines the width of the fitness peak.

In this context we interpret the network weights as corresponding to a specific signalling network (or type of cancer), and investigate the typical properties of such networks with the goal of learning something about how perturbations propagate and interact in genotype-phenotype (GP)-maps.

\begin{figure}[!htb]
\begin{center}
\includegraphics[width=12cm]{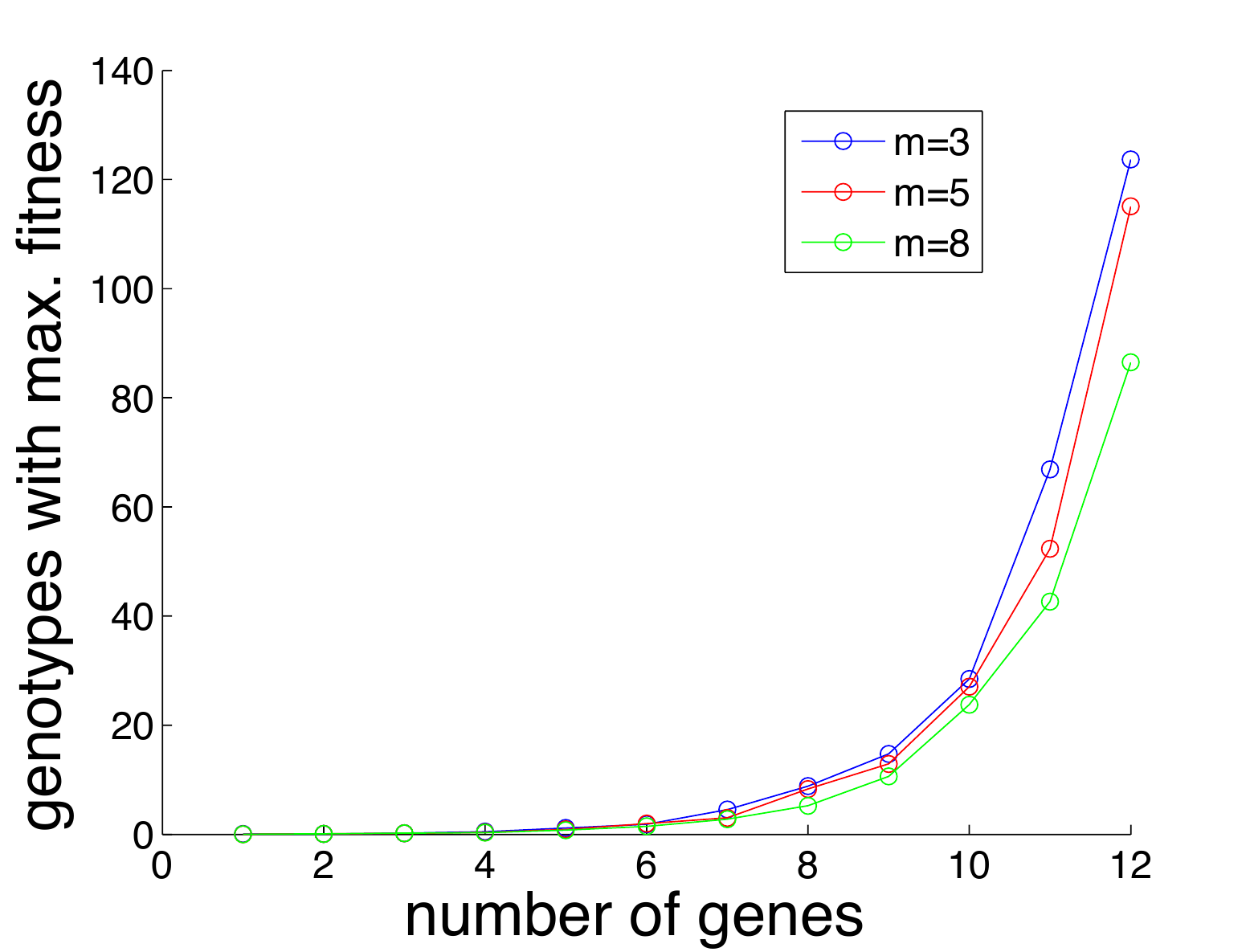}
\caption{\label{fig:fig0}The number of genotypes that map to $F \approx 1$ as a function of the number of genes in the network. The different curves correspond to varying the number of pathways (hidden nodes) in the network. Results are averaged over 100 realisations.} 
\end{center}
\end{figure}

rThe GP-map, and specifically the set of subclones that have a high fitness, depends on the weight matrices ($\mathbf{W}$ and $\mathbf{V}$). In this general framework they remain unspecified, but it is still possible to investigate the typical behaviour of the model by analysing the average behaviour across many network realisations. In figure \ref{fig:fig0} we have made use of this technique to quantify the number of genotypes that map to maximal fitness ($F\approx 1$). This quantity was estimated by creating 100  GP-maps, where the elements of $\mathbf{W}$ and $\mathbf{V}$ were chosen at random from a uniform distribution in the interval $[-1,1]$, and for each network realisation counting the number of genotypes for which $|F-1|<10^{-2}$. From this figure it is clear that the number of genotypes increases exponentially with the number of genes in the network, and shows that even for small $n$ there are many genotypes that have a similair phenotype. This implies that although the fitness landscape (as a function of phenotype) is single-peaked we can still have many distinct genotypes with high fitness \cite{Romero2009}. 


\subsection*{Population dynamics}
To capture the growth of a tumour made up of a heterogeneous  set of subclones and how it responds to the application of drugs, we couple the GP-map to a population dynamics model. 

In the spirit of Beckman et al.\ \cite{Beckman2012}, who also investigated the impact of heterogeneity on different therapies, we assume that tumour growth is unbounded and that tumour heterogeneity is due to differential drug response among different subclones. 
This is of course a gross simplification, but allows us to focus on the relevant aspects of clonal evolution and the acquisition of resistance. The GP-map described above defines the growth rate of each subclone based on its genotype, and we further assume that mutations occur at a rate $\mu$ during cell division. Each subclone can be represented using a bit string of length $n$, e.g. $G=(011010)$, where each binary number corresponds the mutational status, $g_i$, of each gene considered. In terms of mutations we restrict ourselves to one step mutants, which means that the possible mutants produced by a subclone all lie at Hamming distance one from that subclone {(i.e.\ they only differ by one mutation). For generality we allow for back mutations to occur (i.e.\ $1 \rightarrow 0$), despite the fact that they represent an unlikely biological event.} 

Since cell numbers in a tumour typically are very large we model this system using ordinary differential equations and formulate the following equation for the subclone abundances $N_i$ (where $i$ is the decimal equivalent of the binary genotype):

\begin{equation}
\frac{dN_i}{dt} = \left(\alpha \sum_{j=1}^{2^n} Q_{ji}F_j N_j - \delta N_i \right) I(N_i)
\end{equation}
where $\alpha = \log 2$ is the maximal growth rate and $0 < F_j < 1$   is the relative growth rate of genotype $j$. $Q_{ij}$ is the rate at which subclone $i$ produces offspring of type $j$, $\delta=10^{-2}$ is a death rate constant across all clones, and $I(\cdot)$ is an indicator function defined as 
\[I(x) = \left\{
  \begin{array}{lr}
    1 & : x \ge 1\\
    0 & : x < 1
  \end{array}
\right.
\]
which guarantees that fractional cell numbers do not contribute to cell division. {The sum across all subclones might seem counter-intuitive, but please note that $Q_{ij} = \mu =10^{-6}$ only if $i$ and $j$ are neighbours in genotype space, $Q_{ii} = 1-n\mu$ (the net growth rate adjusted for mutant daugther cells lost to other subclones), and that all other entries are zero}. Given the clonal origin of cancer we initiate the model with a single subclone that is one mutational step away from the wildtype ($G=(0000...)$). 

\subsection*{Application of therapy}
Within the model we can implement anti-cancer therapy by considering perturbations to the GP-map at three distinct levels: 
\begin{enumerate}

\item the inhibition of a single mutant protein $i$, which is achieved by setting $g_i = 0$ (reverting to wildtype) in all subclones, or equivalently by removing all links from gene $i$, i.e.\ setting $W_{ij}=0$ for all $j$

\item inhibition of a single pathway $j$, which is achieved by setting $y_j = 0$ in all subclones

\item application of chemotherapy which is implemented by setting the fitness of all subclones $k$ with $F_k > \theta$ to zero, affecting those sublcones that divide at a high rate.

\end{enumerate}
For the two former therapies we also need to specify which gene/pathway to target. This choice will be informed by a virtual biopsy, and we assume the following simple protocol: target a gene that is mutated in the most dominant clone. And for the pathway-level: target the pathway which has the highest average activity in the population. {The tumour is considered detected and hence the treatment starts when $M=\sum_i N_i > 10^9$. For all three therapies we assume that the therapy lasts until the patient dies of excessive tumour burden, which we assume to be $M = 10^{13}$.} 

\section*{Preliminary results}
\begin{figure}[!htb]
\begin{center}
\includegraphics[width=13cm]{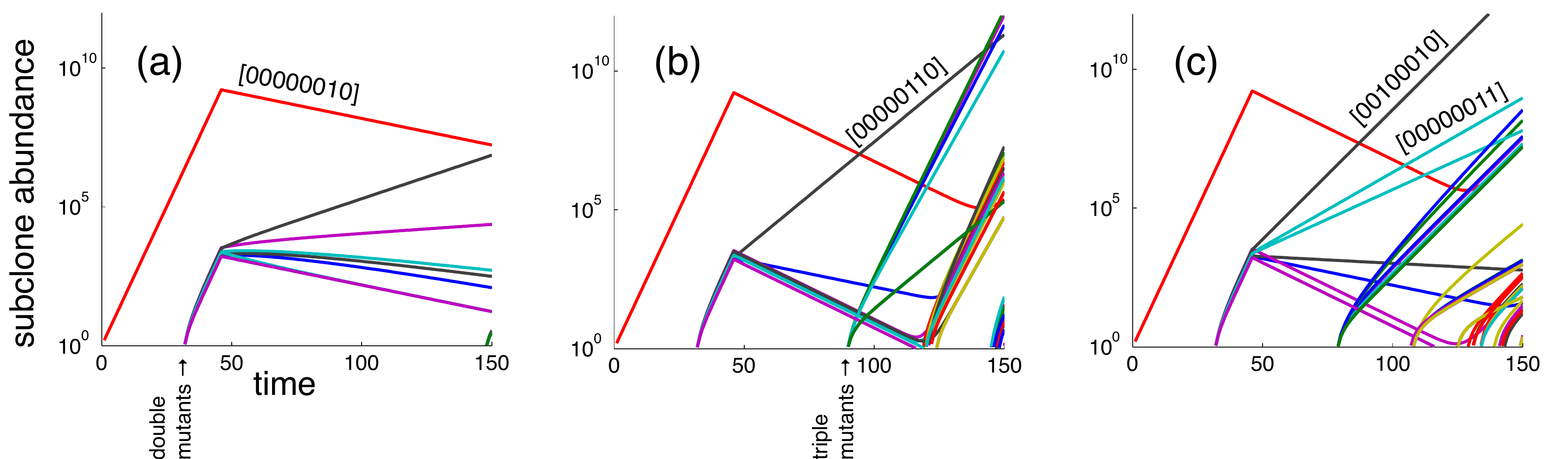}
\caption{\label{fig:clone} {Subclone dynamics under (a) gene targeted therapy (b) pathway therapy and (c) chemotherapy. This clearly illustrates that a single virtual patient responds differently to the three therapies.}} 
\end{center}
\end{figure}

Instead of focusing on a specific GP-map (i.e.\ a specific choice of $\mathbf{W}$ and $\mathbf{V}$) corresponding to a type of cancer (with its specific driver genes and pathways) we investigate the general features of the model and the impact of the therapies described above. We therefore simulate the model for a large number of randomly generated genotype-phenotype maps, where each realisation of can be viewed as a virtual patient with its unique disease state. {Figure \ref{fig:clone} shows the three different therapies applied independently to the same patient, and clearly emphasizes that they have differing impact on the evolutionary dynamics of the tumour. The gene targeted therapy (a) affects all subclones and results in a good response, while the pathway targeted therapy (b) and chemotherapy (c) misses one or more subclones which leads to disease recurrence.}

\begin{figure}[!htb]
\begin{center}
\includegraphics[width=13cm]{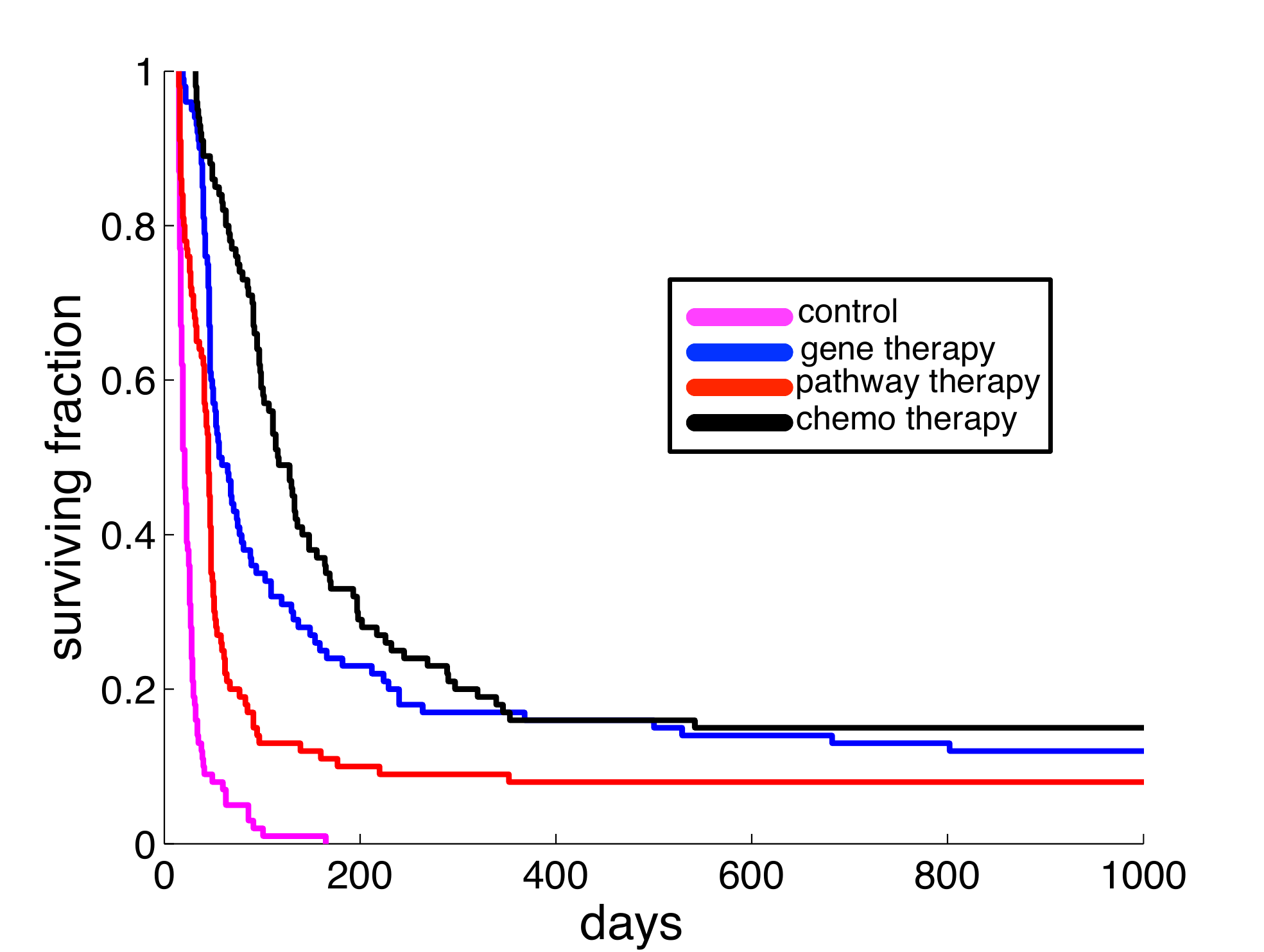}
\caption{\label{fig:survival}Kaplan-Meier survival curves for a virtual patient cohort of 100 patients on which the different therapies were applied. Therapy is applied when the tumour has reached $M=10^{9}$ cells and the patient is assumed to have died when the tumour reaches a size of $M=10^{13}$. All three therapies improve survival compared to the no-therapy control (blue curve).} 
\end{center}
\end{figure}

{The effect of the different therapies on a population level can be seen in figure\ \ref{fig:survival}, which shows the outcome of each therapy in terms of Kaplan-Meier survival curves generated from a cohort of 100 virtual patients. For comparison this plot also contains the untreated case as control (magenta). It is interesting to note that gene-targeted therapy achieves almost the same long-term survival as chemotherapy, but that on shorter time-scales chemotherapy has better survival (the median survival is approximately 50 days for gene therapy, while roughly 100 days for chemotherapy). However, it should be noted that in reality the toxicity of chemotherapeutic drugs is considerably higher than that of targeted therapy, and that the chemotherapy therefore most likely cannot be applied to the extent we have modelled here.}

These preliminary results hint at the potential for this approach in being able to test (and suggest) novel therapeutic strategies not only in terms of mono-therapies but also in terms of combinations of these very different drug  treatments. However, the generality of  this approach means  we are not explicitly considering a known network, making connecting these results to real drugs difficult. In the next section we address this issue by explicitly considering a subset of the kinase pathways.

\section*{3. Pathway activity in normal and cancer cells }
{ In the previous sections the networks we have discussed were feedforward with no direct feedback within each layer (or scale). Whilst a useful simpifying assumption, this approach does not capture the reality that genes can and do regulate other genes and likewise for proteins}. We now therefore describe a recurrent neural network (RNN) model that explicitly allows for regulatory feedback between genes. In RNNs, connections between nodes (genes or proteins) form a directed cycle and therefore can have very different dynamics from the standard feedforward networks. This approach has recently been used to infer genetic networks \cite{Noman2013, Noman2014}. Here, we will use a similar, yet more refined, approach to study key signaling pathways in melanoma (skin cancer).

\begin{figure}[!ht]
\begin{center}
\includegraphics[width = 5in]
{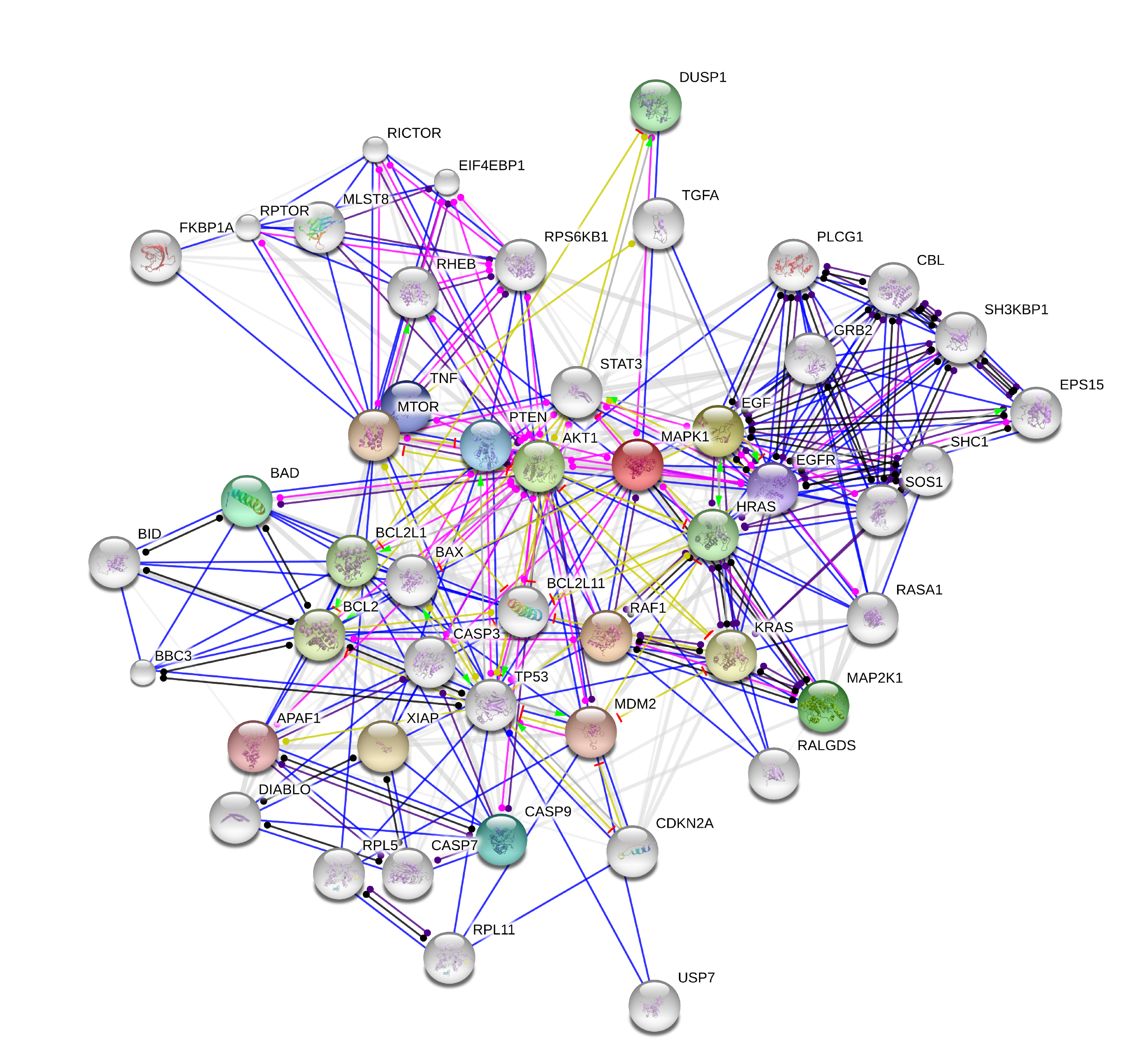}
\end{center}
\caption{
 { Complex signaling network.} A network view generated by the string database (http://string-db.org/) that summarizes the network of predicted associations for key proteins in the MAPK \& AKT/PI3K pathways. The network nodes are proteins, and the edges represent the predicted functional associations.  The colored edges represent function; green-activation, red-inhibition, blue-binding, pink-post-translation, black-reaction, yellow-expression and gray (white)-homology.}
\label{complexF}
\end{figure}

In melanoma cells, tightly regulated signaling networks (fig.\ \ref{complexF}) are significantly altered, leading to uncontrolled melanoma cell proliferation, survival and migration. Of all the signaling pathways, mitogen-activated protein kinase (MAPK) pathways and the PI3K/AKT pathway are considered key players in the regulation of melanoma cell growth and death (see review by Smalley \cite{Smalley:2010dq}). We therefore use our RNN model to describe the MAPK and AKT/PI3K pathways and examine pathway responses and resulting cell phenotypes under different micro-environments (various concentrations of both, growth factors and death signals). We first study the signaling of normal melanocytes and then subsequently evolve this normal network under different micro-environmental constraints to derive a melanoma-signaling pathway.

By constraining the network to specific pathways we can utilize it in many different applications. The most obvious example being to investigate targeted therapies, since these therapies typically target one of the molecules of the MAPK or PI3K/AKT pathways. More specifically, the small molecule BRAF kinase inhibitors have been used for the melanoma patients who harbor the activating BRAF mutation \cite{Flaherty:2010zr,Kim:2010ly,Chapman:2011ys,Hauschild:2012ve} to block the activity of BRAF in the MAPK pathway. Using the model to better understand the signaling response and phenotype of melanoma cells under this therapy will allow us to investigate the underlying mechanisms of drug resistance. The model can also be integrated into multicellular tumor modeling  (for example, our vSkin model \cite{Kim:2013vn}) to investigate the role of signaling perturbations and heterogeneous expression of proteins on melanoma initiation, progression and treatment responses.

\subsection*{Pathway Model}
 
 \begin{figure}[!ht]
\begin{center}
\includegraphics[width = 10cm]{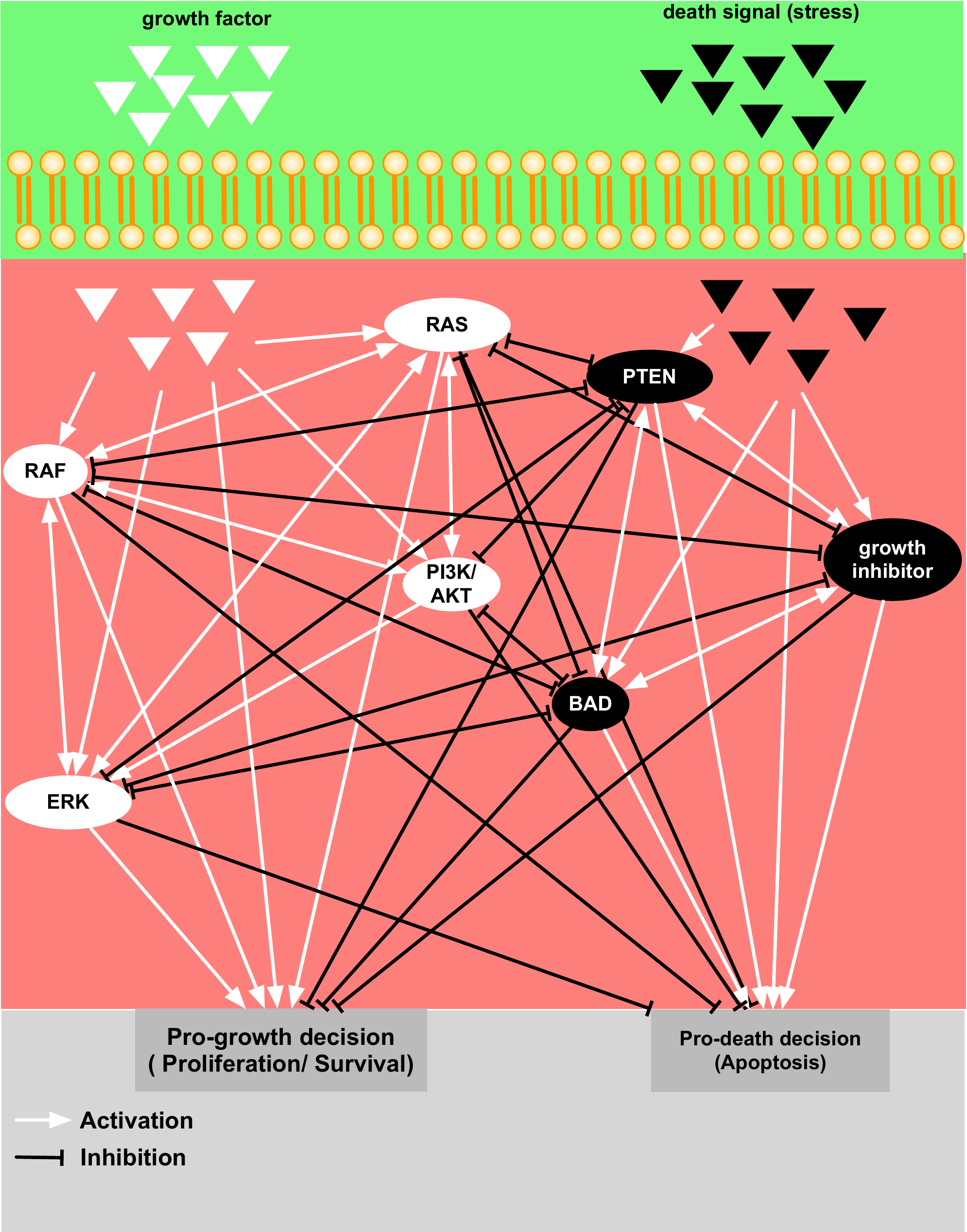}
\end{center}
\caption{
 { Smaller signaling pathway of a melanocyte.} The network is composed of two microenvironmental inputs (pro-growth indicated by white triangle and pro-death represented by black triangles), 7 internal nodes and two cell phenotypes (pro-growth and pro-death). White arrows indicate stimulation or activation, while black arrows represent inhibition or repression.}
\label{sFig}
\end{figure}

Due to computational feasibility, we will only consider a subset of proteins (RAS, RAF, ERK, Growth inhibitor, BAD, AKT/PI3K, and PTEN) from the network (fig.\ \ref{complexF}) and generate a far simpler network (fig.\ \ref{sFig}) that still contains the major network hubs and key targetable proteins. We also include micro-environmental inputs (pro-growth \& pro-death signals) and phenotypic outputs (pro-growth \& pro-death). Only the internal network assumes recurrent interactions, indicated by double direction arrows in figure \ref{sFig}. The expression of each protein is modulated by interactions with its neighbors (adjacency matrix, ${\mathbf A}$ and weight matrix, ${\mathbf{W}}$). We define a prefered network response formalised into a ``goodness function ($G$)" of the network and use a Monte Carlo algorithm to obtain ${\mathbf{W}}$. The algorithm perturbs each element of the weight matrix (${\mathbf{W}}$) and evaluates goodness per iteration. If the perturbed element produces better goodness, we accept the perturbation, otherwise, we discard the change. We iterate this process until the value of the goodness function satisfies a given criteria (see the Appendix for a more detailed discussion of the model).

\subsection*{Derivation of normal cell weight matrix}
In a normal melanocyte, the phenotypic behaviour of the cell (birth or death) is directly regulated by micro-environmental  cues. For a melanocyte that resides in a growth promoting micro-environment (high pro-growth factor but low pro-death factors), it is more likely to reproduce (i.e. a higher probability of cell division). In contrast, a cell living in a growth inhibiting or death promoting micro-environment, should be less likely to divide and will have a higher chance of committing  apoptosis. To obtain a  weight matrix (${\mathbf W}_N$) that models this normal cell behavior, we use the goodness function $G_N$, where subscript $N$ stands for normal. 
\begin{equation}
G_N(W) =  |y_1-y_{n-1}| + |y_2 - y_{n}|,
\end{equation}
where $y_1$ and $y_2$ are given pro-growth and pro-death microenvironmental inputs, respectively. $y_{n-1}$ is a pro-growth output, and $y_{n}$ is a pro-death output. The function $G_N$ minimizes the difference between input and output. It is worth noting that we can change the micro-environmental  conditions $y_{1,2}$ to obtain signaling networks that respond to cells under these micro-environmental changes.

 \begin{figure}[!ht]
\begin{center}
\includegraphics[width = 10cm]{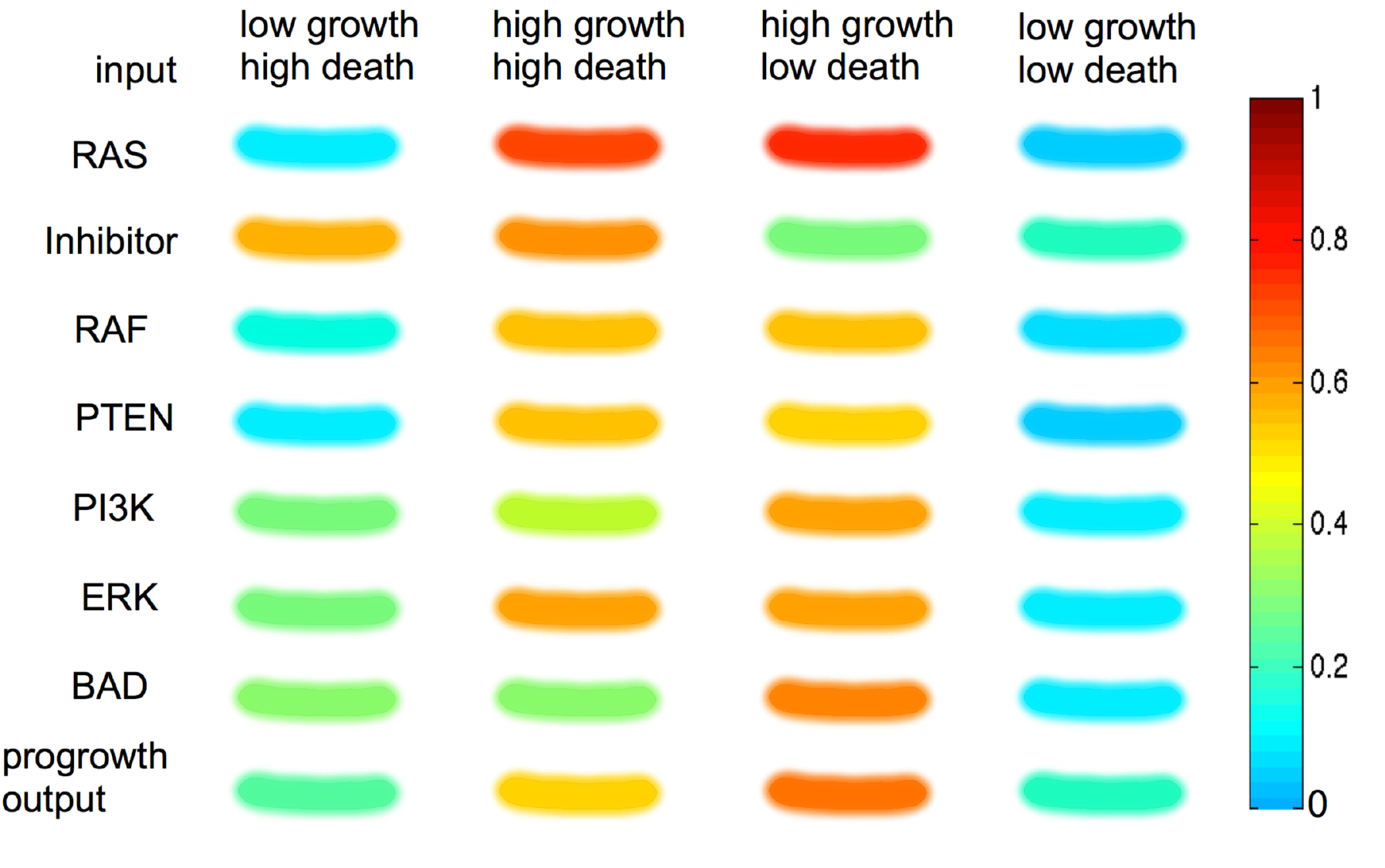}
\end{center}
\caption{
 { {\textit { In Silico}} Gel of a normal melanocyte in different microenvironments.} The protein activity of the pathway is compared in four different micro-environmental  conditions - low-growth \& high-death factor, high growth \& high death factor, high growth \& low death factor, low growth \& low death factor. }
\label{normal}
\end{figure}
The resulting weight matrix (${\mathbf W}_N$) produces a profile of the proteins, shown in figure \ref{normal} in four different micro-environmental  conditions, i) low growth but high death, ii) high growth and high death, iii) high growth but low death, and iv) low growth and low death. Growth factors stimulate the activity of pro-growth proteins (white nodes in fig.\ \ref{sFig}: RAS, RAF, PI3K, and ERK), while pro-death signals increase the activity of pro-death proteins (black nodes in fig.\ \ref{sFig}: Inhibitor, PTEN, BAD) as we might expect but in mixed environments we begin to see more complex dynamics. In  micro-environment (i), the expression of  inhibitor is particularly high while all other protein expressions are quite low. The resulting pro-growth output is quite low (0.3), which  means that a melanocyte in (i) may be less likely to reproduce. In micro-environment (ii), the expressions of all proteins increase and the pathway produces a higher pro-growth output (0.5). The expression is even more increased and produces even higher output (0.6) in condition (iii). Lastly, micro-environment (iv) decreases protein expression, pushing cells toward an inactive state. The relative changes in progrowth output that the normal melanocyte displays in different contexts, gives us some confidence that the network is an appropriate description of a normal melanocyte. Importantly, these results also highlight the diversity of protein expression that the same cell can exhibit purely as a result of different micro-environmental stimulus.

\subsection*{Evolution of melanoma cell lines}
In order to generate a collection of melanoma cell lines we mutate our normal melanocyte network and let it evolve under three different micro-environmental conditions to simulate different selection pressures. More specifically, we evolve the normal weight matrix (${\mathbf W}_N$) derived in the above section, to produce a weight matrix for melanoma cells. We introduce mutations by changing each element of the weight matrix (${\mathbf W}_N$) in a random manner and use the pro-growth output as a goodness function $G_M \ (G_M = y_{n-1})$, where subscript $M$ stands for mutation. If a mutated network produces a greater $G_M$ value (i.e.\ it is more likely to divide), the mutation is accepted. Otherwise, the mutation is discarded. The weight matrix ${\mathbf W}_N$ is evolved in three different conditions, i) a cyclic microenvironment, ii) a pro-death microenvironment (low growth factor \& high death factor), and iii) a pro-growth microenvironment (high growth factor \& low death factor) to generate three different melanoma cell lines. Cell line 1 is evolved in condition (i), cell line 2 in (ii), and cell line 3 is evolved in the condition (iii). 


 \begin{figure}[!ht]
\begin{center}
\includegraphics[width = 10cm]{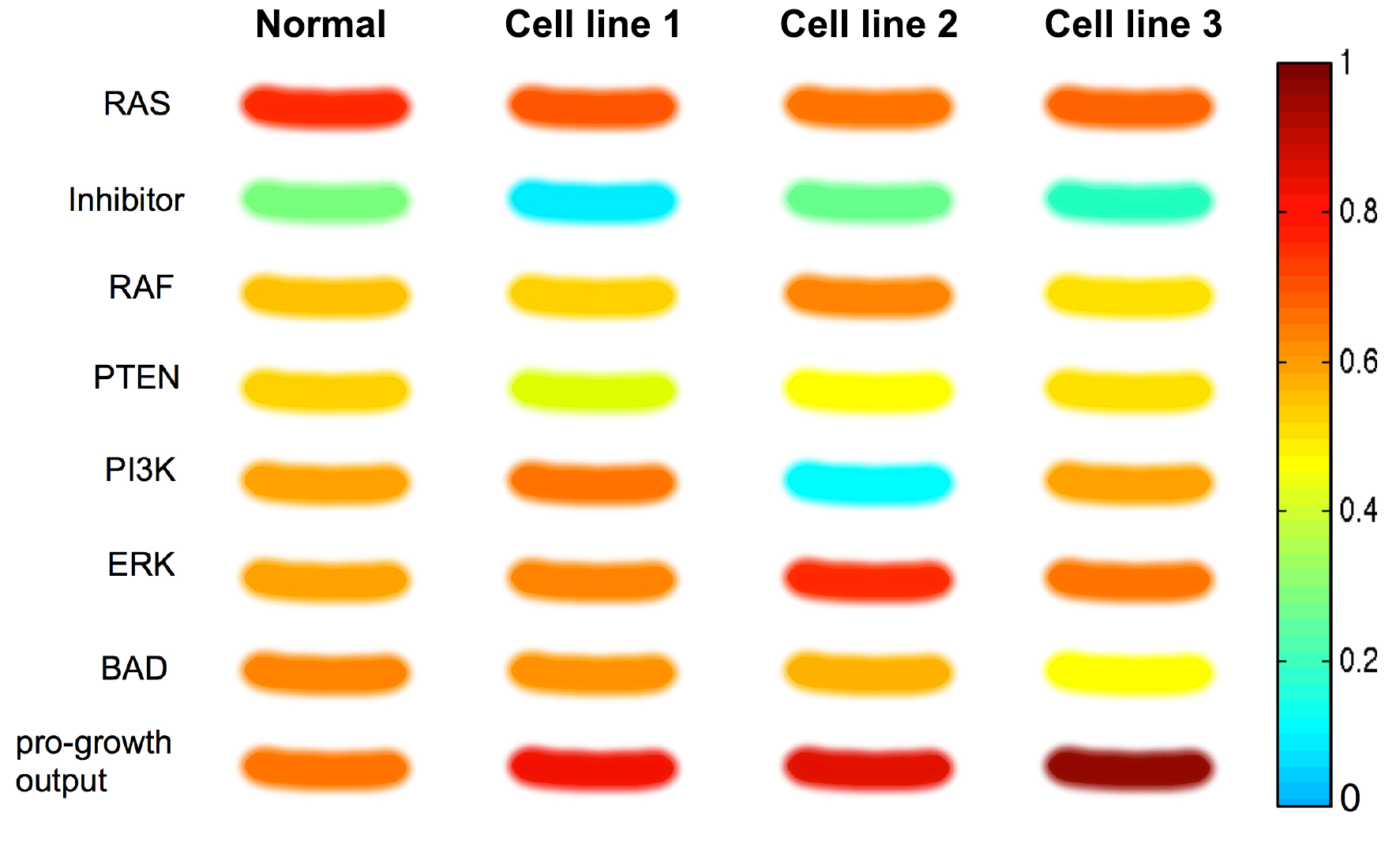}
\end{center}
\caption{{ {\textit { In Silico} }Gel of a normal melanocyte and cancerous cell lines.} The protein expression and pro-growth output of a normal melanocyte and three cancer cell lines (1-3) in microenvironment condition (iii), the high growth factor \& low death factor condition.
}
\label{Cancer}
\end{figure}

We compare the typical protein expression of our normal ({\it in silico}) melanocyte and three evolved cell lines (fig. \ref{Cancer}) in the high growth factor \& low-death factor (typical {\it in vitro} culture conditions) microenvironment. Cell line 1 has a significantly lower than normal expression of the inhibitor protein (cyan) which suggests that it may harbour an inactivating mutation. Cell line 2, on the other hand, may have activating mutations in RAF and ERK as well as an inactivating mutation in PI3K. The expression of cell line 3 is only slightly different from that of a normal cell. Since cell line 3 was evolved in the grow-promoting environment, only a slight change in the network was sufficient to satisfy the given criteria (i.e., high pro-growth output). Due to the complexity of even this simple network, these cell lines will only truly reveal their full potential when exposed to many different environments and compared to normal. Of course, the critical environment that we are most interested in, is drug.

\section*{Discussion}
In this paper we have reviewed the application of neural networks as models for the relationship between the micro-environment, genotype, pathway activity and cellular phenotype in tumour cells. Describing all these components simultaneously within one model would require a model whose complexity most likely would render it {incomprehensible and hence useless}, and and we have therefore described three separate approaches that focused on different subsystems of the problem, hence breaking down its complexity to a manageable level. 

The neural network methodology has the advantage that it has the capability of capturing  dynamics on two different time scales, that of pathway activity, with gene activity being influenced by micro-environmental factors and the state of other genes, and the time scale of evolutionary change, where the expressed phenotypes are selected for and the respective sublcones might alter the micro-environment in a feedback loop (see fig.\ \ref{fig:schematic}). This makes these models suitable tools for investigating and interrogating the evolutionary dynamics of tumour growth. However, we do not expect that a single model can encompass all aspects of this process, but rather that the problem needs to be attacked as it was presented in this article, in a piecemeal fashion. The first approach we discussed focused on the feedback between environment and phenotype, the second case on how the mapping from genotype to phenotype influences drug efficacy, and the last case investigated the impact of feedback of proteins on oneanother and how this alters the mapping from environment to phenotype.

The latter approach gave some insight into the architecture of melanocyte signalling networks, but in terms of identifying the structure of genetic regulatory networks, there is a plethora of methods that address this problem with a high degree of statistical rigour, e.g.\ dynamic Bayesian networks \cite{Hill2012}, mutual information \cite{Margolin2006} or regularised least squares \cite{Jornsten2011}. However, the underlying assumptions about the dynamics of gene regulation are often highly restrictive (e.g.\ steady-state, linearisation), and do not lend themselves well to the time- and context-dependent dynamics of the kind described in the last section. In a sense it could be said that the neural network approach presents a trade-off between network inference and the ability to model the evolutionary dynamics of the system. { As we uncover more details about the rich spatial heterogeneity in cancer, understanding the role of context in driving selection as well as the role of the tumour in creating context is becoming increasingly important.} 

An obvious drawback for the neural network approach is the high level of abstraction it requires. For example the pathways present in the feed-forward network in the first section (see fig.\ \ref{fig:cellnet}) do not correspond to any particular pathways, but rather represent collections of pathways that influence the respective output nodes. Of course this can be made more precise, as in last section, where the expression levels of actual proteins are modelled. Similarly, in section 2 the mapping from genotype to phenotype is highly non-specific, which, although it makes the model harder to validate, has the possible advantage that the results derived from it can be applied to a wide range of different tumour types and targeted therapies. Naturally the model could also be made more specific and detailed if a specific type of cancer was to be modelled. 

{The prospect of going even further and tailoring these models to specific patients is alluring, especially in the current climate of personalised medicine. However, before that becomes a reality we need to develop models that are refined to such a degree that patient data (mutational status, subclone abundance, microenvironmental status etc.) becomes meaningful in the context of the model. This raises another issue, in the clinic the actual measurements that are taken from patients largely revolve around single timepoint whole organ imaging, blood and possibly tissue samples. Making a clinically relevant model that incorporates even the simplest of the models presented in this review would mean a paradigm shift in how and what we measure in a patient. This of course does not mean we should not try to develop the methods before they become truly translatable.}

However, in order to achieve this goal, we do not just need more sophisticated models, but more importantly a closer connection between models and experiment (and ultimately between experiment and clinic). In terms of the models discussed here this amounts to a better understanding of how micro-environmental variables affect both pathway activity and cellular phenotypes. Quantifying the latter poses a significant challenge, since the phenotypes observed \textit{in vivo} are often difficult to reproduce and quantify in the lab. This is starting to change, e.g.\ with the development of organotypic cultures that represent a caricature of the real system containing not only cancer cells but also stromal cells \cite{Oyama:2007vn,Brohem:2011ys,Vorsmann:2013zr} and \textit{ex vivo} assays \cite{Farin2006,Vaira:2010ly} that can keep a slice of real tissue alive  for some time. In addition, we also need a better understanding of the evolutionary pressures affecting tumour growth, and here a careful comparison of model prediction and experimental interrogation can help advance our knowledge of the complex evolving ecological system that is cancer.

\section*{Acknowledgements}
We gratefully acknowledge the NIH/NCI support from U01 CA151924, the ICBP (U54 CA113007) and PSOC (U54 CA143970) programs.

\section*{Appendix}

\subsection*{Network parametrisation for agent-based model}
\begin{figure}[!htb]
\begin{center}
\includegraphics[width=13cm]{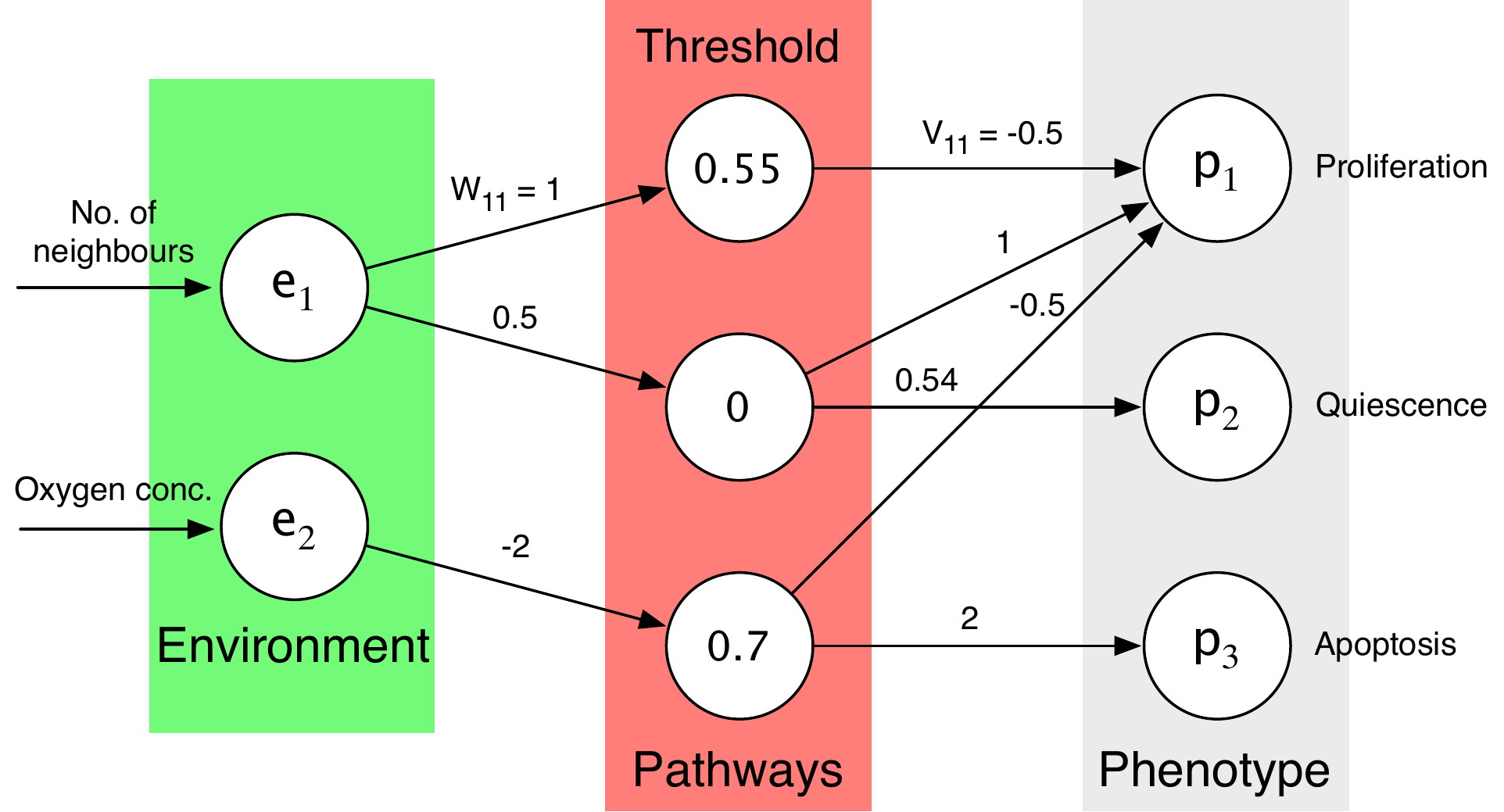}
\caption{\label{fig:cancernet} {The layout of the initial network used in the agent-based model of tumour growth. The input to the network is the number of neighbours of the cell and the local oxygen concentration. The output of the network corresponds to proliferation (P), quiescence (Q) and apotosis (A), out of which the strongest is chosen.}} 
\end{center}
\end{figure}

In the most basic setting of our model the only input to the network is the number of neighbours of the cell and the local oxygen concentration. The reason for this choice is that cancer cells often show weaker response to hypoxia-induced apoptosis \citep{apop} (programmed cell death) and that they tend to adhere less to their neighbours \citep{adhesion}. This implies that the input vector $\mathbf{e}$ will have two components, $\mathbf{e} = (n(\vec x,t),c(\vec x,t))$, where $n(\vec x,t)$ is the number of neighbours and $c(\vec x,t)$ the oxygen concentration. The number of neighbours determines if the cell will proliferate (if $n(\vec x,t) > 3$) or become quiescent (if $n(\vec x,t) \leq 3$), while the oxygen concentration influences the apoptotic response. If the oxygen concentration falls below a certain threshold $c_{ap}$ the apoptosis node is activated and the cell dies. An initial network that fulfilled the above specifications was created and used as a "seed" in every simulation (see fig. \ref{fig:cancernet}). For further details we refer the reader to \cite{Gerlee2009b}.

\subsection*{Pathway activity in normal and cancer cells}

In the pathway model, the activities of proteins in the signaling network are represented by a vector ${\mathbf{y}} \in \mathbb{R}^n$. The first two elements of $ {\mathbf{y}}$ represent pro-growth ($y_1$) and pro-death ($y_2$) inputs and the values are given as external conditions. We derive the rest of elements by using an adjacent matrix ${\mathbf{A}}$ that models the connectivity of network and a weight matrix ${\mathbf{W}}\ (W_{ij} \geq 0)$ that models the strength of each connection \cite{Noman2013, Noman2014,Weaver:1999fk}. The adjacency matrix ${\mathbf{A}}$ is defined as

\[ 
A_{ij}= \left\{ \begin{array}{ll}
         1 & \mbox{if an input node $j$ activates an output node $i$, }\\
         -1 &  \mbox{if an input node $j$ inhibits an output node $i$, }\\
         0 & \mbox{otherwise (no relation)}.\end{array} 
        \right. 
 \]
 
To obtain a weigth matrix, we use Monte Carlo Algorithm. A weight matrix is initialized with non-negative random numbers. We use the elementwise product of the connectivity matrix ${\mathbf{A}}$ and an weight matrix ${\mathbf{W}}$ to obtain interaction matrix, ${\mathbf{V}}\ (V_{ij} = A_{ij} \cdot W_{ij})$. The effect of $j$ on $i$ is simply modeled as a product of the expression level of  influence, $y_j V_{ij}$. The total influences of the neighbor of a node $i$ is then the sum of all of its neighbors
\begin{equation}
F_i = \sum_{j=1}^{n} V_{ij} y_j, \ \ i = 1,2,...n
\label{induction}
 \end{equation}

We update the protein activity levels by solving the ordinary differential equation

\begin{equation}
\frac{dy_i}{dt} = T\left( \sum_{j=1}^{n} V_{ij} y_j \right) - \lambda y_i(t), \ i = 3,4,... n,
\label{eqODE}
\end{equation}
where $y_i (0) = 0, i = 3,4,...n$ and $\lambda$ is the decay rate of the proteins. The function $T$ is a transfer function $\left(T(x) =  \frac{1}{1+e^{- \beta x}} \right)$. We use this transfer function to account for saturation effects. The solution of the equation (\ref{eqODE}), $y_i$ is final protein expression of protein $i$ in the network. Note that $y_{1,2}$ are given as microenvironmental (external) conditions and we do not update these nodes using the above equation. 

Next, we use updated protein activity levels to evaluate the pre-defined goodness function ($G_N$ or $G_M$). We then randomly select an element of weight matrix and perturb it. We then use equation~(\ref{induction}-\ref{eqODE}) to obtain updated protein activities. We use these updated levels to evaluate goodness function. We compare this updated goodness function value to previous one, and accept the change of the weight element if this new value is closer to pre-defined (desired) criteira of goodness value. We iterate this process until convergence.


\bibliography{SICB}
\bibliographystyle{vancouver}

\end{document}